# Photonic Neural Network Fabricated on Thin Film Lithium Niobate for High-Fidelity and Power-Efficient Matrix Computation

Yong Zheng[1,2,†], Rongbo Wu[2,8,†], Yuan Ren[1,2], Rui Bao[1,2], Jian Liu[1,2], Yu Ma[3], Min Wang[2] and Ya Cheng[1,2,3,4,5,6,7]*

[1]*State Key Laboratory of Precision Spectroscopy, East China Normal University, Shanghai 200062, China*

[2]*The Extreme Optoelectromechanics Laboratory (XXL), School of Physics and Electronic Science, East China Normal University, Shanghai 200241, China*

[3]*State Key Laboratory of High Field Laser Physics and CAS Center for Excellence in Ultra-intense Laser Science, Shanghai Institute of Optics and Fine Mechanics (SIOM), Chinese Academy of Sciences (CAS), Shanghai 201800, China*

[4]*Collaborative Innovation Center of Extreme Optics, Shanxi University, Taiyuan 030006, China.*

[5]*Collaborative Innovation Center of Light Manipulations and Applications, Shandong Normal University, Jinan 250358, China*

[6]*Shanghai Research Center for Quantum Sciences, Shanghai 201315, China*

[7]*Hefei National Laboratory, Hefei 230088, China*

[†]*The authors contribute equally*

[8]*rbwu@phy.ecnu.edu.cn*

[*]*ya.cheng@siom.ac.cn*

**Abstract**

Photonic neural networks (PNNs) have emerged as a promising platform to address the energy consumption issue that comes with the advancement of artificial intelligence technology, and thin film lithium niobate (TFLN) offers an attractive solution as a material platform mainly for its combined characteristics of low optical loss and large electro-optic (EO) coefficients. Here, we present the first implementation of an EO tunable PNN based on the TFLN platform. Our device features ultra-high fidelity, high computation speed, and exceptional power efficiency. We benchmark the performance of our device with several deep learning missions including in-situ training of Circle and Moons nonlinear datasets classification, Iris flower species recognition, and handwriting digits recognition. Our work paves the way for sustainable up-scaling of high-speed, energy-efficient PNNs.

Artificial neural networks (ANNs) that power deep learning algorithms to mimic the human brain have progressed explosively over the last decade [1,2], leading to disruptive artificial intelligence (AI) technologies ranging from natural language processing to self-driving cars [3,4]. Owing to the increasing complexity of the tasks to accomplish with ANNs, the energy required to train and infer ANNs doubles every half year [5]. This brings about the crucial need for developing innovative energy-efficient hardware for ANNs to support the sustainable growth of AI in the future. In recent years, photonic neural networks (PNNs) [6–18] based on photonic meshes consisting of programmable Mach-Zehnder interferometers (MZIs) and phase shifters (PSs) [19–21] have emerged as a promising platform to address the power consumption problem. By sending input light signals with predesignated amplitudes and phases through the MZI meshes and measuring the changes in the output signals, PNNs accelerate the unitary matrix-vector multiplication (MVM), the most expensive operation in machine learning tasks, by orders of magnitude in its speed. In addition, PNNs hold the potential to be operated with extremely low energy consumption as far as the systems feature low optical loss as well as high phase-tuning efficiency of the MZI meshes.

However, the idea of PNNs is difficult to implement despite its apparent advantages as mentioned above. This is due to the formidable challenges in realizing high-fidelity, large-scale integrated photonic circuits. The field of large-scale PNNs did not really take off until silicon photonics technology matured in the past ten years. After YC Shen demonstrated the first PNN on silicon-on-insulator (SOI) [10], substantial progress has been made including a PNN capable of performing complex-valued arithmetic [18], a PNN with programmable nonlinear optical function units and coherent receivers that can execute inference and in-situ training [14], and a PNN capable of performing in-situ back-propagation [11]. Notably, these PNN systems almost exclusively use thermo-optic (TO) tuning to achieve phase modulation. When the PNNs scale up, severe challenges such as high TO crosstalk, high power consumption for phase tuning, low reconfiguration rate of MZI meshes, and increased on-chip optical loss all come together. To tackle the issues, alternative material platforms such as silicon nitride [15,22] and thin film lithium niobate (TFLN) [23] have been extensively investigated with proof-of-concept demonstrations. The unique advantage of using TFLN as the platform for PNNs is that lithium niobate has a large electro-optic (EO) coefficient, which allows fast and efficient phase tuning with extremely low power consumption. This advantage becomes more prominent when the scale of PNNs expands to comprise large number of MZIs. In combination with the ultra-low propagation loss at the level of ~3 dB/m in the LN waveguides [24,25], TFLN becomes an attractive platform to demonstrate high-fidelity low power-consumption PNNs for deep learning applications. Our previous work demonstrates a TFLN 4×4 programmable linear PIC to perform a general special unitary group of degree 4 (SU(4)) with a low power consumption of 1.5 mW when operated at 100 MHz modulation, laying the foundation of TFLN-based large-scale PNNs [26].

In this work, we report the first demonstration of an integrated, coherent PNN fabricated on the TFLN platform, which can implement any 4×4 MVM by embedding an SU(4) in an 6×6 triangle network of MZIs. Once the networks have been trained, such TFLN-based PNN devices utilizing EO effect for phase turning hardly require any power to maintain the configurations of PSs during the forward-propagation computing. Thus, we name them "Zero Energy-consumption Neural-network", abbreviated as ZEN. The current work describes the first generation of ZEN, namely, ZEN-1. To maintain high fidelity of ZEN-1, we use a cascaded MZI architecture to ensure a high extinction ratio of approximately -40 dB. To achieve high computation speed and high energy efficiency, we carefully design impedance-matched co-planar waveguide (CPW) electrodes. To characterize ZEN-1, we benchmark it with multiple AI missions, including in-situ training of Circle and Moon nonlinear dataset classification, in situ training of Iris species recognition, and handwriting digits recognition. We also can prove that the power consumption for a single operation of TFLN-based PNN devices continuously decrease as they scale up. To the best of our knowledge, ZEN-1 is the largest TFLN-based photonic integrated circuit (PIC) ever reported, paving the avenue to the sustainable scaling up of high-speed and energy-efficient ANNs.

A typical hybrid PNN architecture consists of cascaded programmable unitary MVM operations assisted by electronic hardware to provide digital nonlinear functions as well as the input and output layers deployed on both ends of the device, as shown in Fig. 1(a). Our architecture implements the nonlinear functions and the input/output layers through digital software executed on a digital system. The unitary MVM operations are carried out in a photonic fashion using ZEN-1, which comprises two 1×5 photonic meshes on each end of a programmable 4×4 triangle MZI mesh, as schematically shown in Fig. 1(b). During the in-situ training and inference, the initial input light signals produced by a narrow line-width laser was coupled from the bottom port of ZEN-1 (see, upper panel of Fig. 1(b)) using an optical fiber array. The 1x5 cascaded MZI mash marked in red at the leftmost part is referred to as a "generator" [27], which splits light into different waveguides and encodes the input data $x$ into the amplitudes and phases of the light beams propagating in the different waveguides, and meanwhile produces a reference light beam. The programmable 4×4 MZI mesh marked in green in the middle part, as shown in Fig. 1 (b), transforms the light signals produced by the generator to an array of the output light signals $y$ according to $y = Ux$, where $U$ presents an arbitrary SU(4) matrix determined by the configurations of PSs in the 4×4 MZI mesh. The output light signals and the reference light beam were directed to the "analyzer" [27] at the rightmost end of the PNN, as marked in blue in Fig. 1 (b). The analyzer is a series of coherent detectors consisting of a PS and an MZI with photodiodes connected to both output ports. It can extract the amplitudes and phases of the output signals. More details of the working principle of the generator and analyzers can be found in the Supplementary Materials.

In our PNN architecture, the MZIs in the 4×4 mesh are alternative sequences of 3 three directional couplers (DCs) and 3 PSs, as illustrated in the green dashed box in Fig. 1(b). The configuration is chosen as it requires minimum number of tunable PSs to function as a "perfect MZI" with infinitely high ER, from a theoretical point of view, despite the inevitable fabrication errors. The MZIs in the generator and analyzer are constructed with 4 DCs and 4 PSs alternatively arranged to perform as the "perfect MZIs," as illustrated in the red and blue dashed boxes in Fig. 1(b). All the PSs in ZEN-1 are electronically tuned utilizing the EO effect with the horizontal Z-axis of TFLN oriented perpendicularly to the waveguides. This allows the ground-signal-ground

(GSG) electrodes to apply an equal but opposite electric field in the two arms of each PS when the electric voltage is applied, resulting more efficiently in a phase difference in the transmitted light beams. The internal and external PSs in these MZIs are driven by the impedance-matched CPW electrodes for achieving high-speed phase modulation.

The photograph of the ZEN-1 captured with a digital camera is shown in Fig. 2 (a), which is fabricated on a 4-inch X-cut TFLN wafer using PLACE technology. More fabrication details can be found in the Supplementary Materials. As can be seen in the zoom-in microscope image, each PS is 5 mm long, and the total path-length of the waveguides in ZEN-1 is approximately 20 cm. The bends in the upstream waveguides are intentionally designed to filter out the undesirable TM mode, resulting in a high extinction ratio (ER) of the MZIs and in turn the high fidelity of the MVM operation.

To ensure high fidelity, all PSs in ZEN-1 were calibrated for phase offset caused by fabrication fluctuations before device characterization. The details of the calibration process can be found in the Supplementary Materials. Fig. 3(a) shows the measured EO response of a representative MZI in ZEN-1, evidencing a half-wave voltage ($V_\pi$) of 7.33 V and ultra-high ERs of -41.2 and -38.4 dB for the Bar and Cross routes, respectively. To verify the high-speed EO response of ZEN-1, we measured the optical transmission of an MZI as a function of time when applying a 0.1 GHz square-wave voltage to the CPW electrodes of the same MZI through impedance-matched coplanar probes. The result is presented in Fig. 3 (b) and indicates a rise/fall time of 180 ps for the MZI under test. The rise/fall time is mainly limited by the bandwidth of the arbitrary waveform generator (AWG) used to produce the electronic signal, which is below 5 GHz. To further characterize the radio frequency (RF) response of our device, we measured the EO S21 curve of the MZI with a vector network analyzer (VNA). The result, as depicted in Fig. 3(c), shows a 3-dB bandwidth over 20 GHz. The relatively high $V_\pi$ and low bandwidth of our device compare to the most advanced TFLN modulators are due to the fact that our electrodes are fabricated on top of the 1.5-um-thick cladding layer of $SiO_2$ coated on the optical waveguides fabricated on TFLN substrate. This configuration avoids the interference between the electrodes and waveguides whilst simplifies the fabrication process as compared with the the double-layer electrode configuration for higher tuning efficiency and bandwidth. We employed 1 kHz synchronous square wave signals carrying bias voltages for all PSs to address the bias drift problem of LN [28].

The essential idea behind the PNN is to solve the two ultimate challenges of ANN, i.e., the enormous demands on energy consumption and computation power. The computation speed of PNNs can be characterized by calculating the number of operations taken per second (OPS) as $R = 2mN^2f$ [10], where $m$ is the number of layers, $N$ is the number of modes, and $f$ denotes the system bandwidth. In ZEN-1, $m$ equals 1, $N$ equals 4, and $f$ equals 20 GHz. Thus, the computation speed can be calculated as 0.64 tera-operations per second (TOPS). Unlike its counterparts relying on TO turning to achieve phase modulation and inevitably continuously consume energy, ZEN-1 requires hardly any power to maintain the trained configurations of the MZI mesh. The power consumption occurs only when the applied voltages are changed and can be estimated by considering the charging and discharging of a capacitor consisting of the GSG electrodes and TFLN optical waveguides, which can be estimated as $\Delta E = CV^2/2$ [29], where $C$ is the capacitance of PSs and $V$ is the charging voltage. In our device, $C$ is calculated as 15 fF by simulating the electrical field profiles of the GSG electrodes. The charging voltages vary from 0 to $V_\pi$ randomly during the calculation. Thus, the average power consumption of a single charging/discharging process can be estimated as 134 fJ according to the following function:

$$\Delta E_{ave} = \lim_{N \to \infty} \sum_{n=1}^{N} C \frac{n^2 V_\pi^2}{2N^3} = \frac{CV_\pi^2}{6}. \tag{1}$$

The total power consumption for a TFLN-based $m$-layer PNN employing EO turning with a mode number of $N$ can be expressed as follow:

$$E_{total} = \frac{CV_\pi^2 Nf}{3} + m\frac{CV_\pi^2 N(N-1)f_r}{2}, \tag{2}$$

Where $f_r$ = 1 kHz denotes the refresh rate applied for all PSs to address the bias drift problem of LN. In ZEN-1, the charging/discharging processes mainly take place in the 8 CPW electrodes in the generator, and $2N^2 = 32$ operations were taken in a single charging/discharging cycle. Thus, the power efficiency of the ZEN-1 can be estimated as 33.5 fJ/OP.

Furthermore, we highlight the advantages of using EO turning instead of TO turning by evaluating the growth of energy consumption with the scaling up of PNN. Since $f_r$ is much smaller than $f$, the second term in formula (2) is much smaller than the first when $mN \ll f/f_r \sim 10^7$, this leads to an almost linearly dependence of the total power consumption on $N$ for a TFLN-based PNNs like ZEN. In contrast, for a PNN with the same architecture while utilizing TO turning to achieve phase modulation, of which typically the average power consumption of a single PS reaches $E_{TO} \sim 10\ mW$ [14], the total on-chip power consumption can be evaluated as:

$$E_{total} = mN(N-1)E_{TO}, \tag{3}$$

which leads to a quadratically dependence of the total power consumption on $N$. Fig. 3 (d) shows the plot of the total on-chip power consumption of PNNs that use EO/TO turning with *m*=1 and *m*=10 versus *N*, reflecting a huge difference in the two situations. The power efficiency of a PNN device can be characterized by dividing the total power consumption by the number of operations taken per second as $E_{total}/(2MN^2 f)$. We present the power efficiency plots achieved with the two technical approaches in Fig. 3 (h). For the TO turning devices, we assume a bandwidth of the generator of 20 GHz, which is apparently impossible for TO turning but may be achievable using silicon-free carrier modulators. Clearly, the power efficiency of the TO turning-based PNNs remains a fixed value as *N* increases, whilst the power consumption per operation of TFLN-based PNNs continuously drops as *N* increases. The nice power scaling law provides the strongest incentive for developing the very large-scale PICs on the TFLN platform, including the future generations of ZENs.

The overall performance of PNNs can be further characterized by examining their fidelities $\mathcal{F}_i = Tr(|U_i^\dagger \cdot U_{exp}|)/N$ [14], where $U_i^\dagger$ is the conjugate transpose of an arbitrary target unitary matrix $U_i$, $U_{exp}$ is the measured transformation matrix of the PNN, and *N* denotes the mode number. To characterize the fidelity of ZEN-1, instead of using a decomposing process that requires offline computations on a digital computer to determine the voltage configurations of all PSs [30], we implemented a more straightforward approach by conducting 1000 in-situ training with a cost function of

$\mathcal{L} = 1/\mathcal{F}_i$ on our device aiming at 1000 randomly generated unitary matrices. We used a stochastic optimization method that performs gradient descent on average and then converges to a local minimum [14]. More details can be found in Supplemental Materials.

Fig. 3 (e) shows the measured fidelities of ZEN-1 aiming at 1000 randomly generated unitary matrices during in-situ training. In Fig. 3 (f), the 1000 fidelities after 1, 25, 50, and 500 iterations are shown, and the histograms are displayed in Fig. 3 (g) and (h). This indicates that our device achieves an average fidelity of $\langle \mathcal{F} \rangle = 0.985 \pm 0.003$, which is compatible with the highest reported fidelity ($\langle \mathcal{F} \rangle = 0.987 \pm 0.007$) [14] for a programmable photonic matrix processor.

To test the effectiveness of on-chip training, we trained our chip to classify two labeled noisy synthetic datasets - the Circle and the Moons. The objective of the classification was to assign a label of either -1 or 1 according to the spatial location of the data points. The three-layer architecture we used to accomplish this mission is shown in Fig. 4 (a). We began with generating 500 labeled samples for each of the Circle and the Moons dataset using Scikit-Learn [31]. Afterwards, we batched a training set of 400 samples into our system, and the input data $x = (x, y)$ was then encoded to light intensities of the four input waveguides of the SU(4) mesh according to the following formula $|x|^2 + |y|^2 + 2|p|^2 = 1$, such that all the samples had the same total input power as a four-port vector $(x, y, p, p)$. At the output of each hidden layer, we implemented a nonlinear function of $f(x) = |x|^2$ by simply measuring the output optical power from the four outputs. At the end of the final hidden layer, we implemented an output layer on a digital computer with a transmission matrix $T_{out} = (1, 1, -1, -1)$. The entire structure can be summarized as the model given below:

$$z = (1 \quad 1 \quad -1 \quad -1) \left| U^{(3)} \left| U^{(2)} |U^{(1)} x|^2 \right|^2 \right|^2. \tag{4}$$

Each input was classified into red or blue (1 or -1, respectively) based on whether the output satisfied $z>0$. A binary cross-entropy cost function was utilized throughout the training. We preserved 100 samples that the model had never seen as the test set to avoid overfitting. Fig. 4 (b) and (c) show the decision maps of the two datasets after

2000 iterations obtained by inputting a mesh of samples ($x, y \in (-2.5, 2.5); \Delta x, \Delta y = 0.04$) into the model and calculating their outputs $z$. Their training curves are shown in Fig. 4 (d) and (e), indicating the train/test accuracies of 96.25%/99.00% and 94.50/92.00% for the Circle and the Moons datasets, respectively. We also demonstrated an in-situ training for the classification of Iris flowers and ran all 70000 inferences of MNIST handwritten digits reorganization, as presented in Supplemental Material.

To conclude, we demonstrate an integrated coherent PNN constructed on the TFLN platform, which is named ZEN-1, and successfully train ZEN-1 for performing AI tasks. In total, ZEN-1 comprises 54 individually tunable EO PSs monolithically integrated into one chip, showing an on-chip loss of ~3.1 dB, a computation speed of 0.64 TOPS, energy efficiency of 33.5 fJ/OP and a fidelity of 98.5%. With its high performance, ZEN-1 provides convincing evidence of the sustainable scalability of TFLN-based PNNs, as there is still much room to expand the PNN without suffering the pain of growing optical loss and power consumption. Moreover, the PLACE fabrication technique intrinsically supports the patterning of large TFLN wafers (currently up to 6-inch wafers limited by the commercial suppliers) with high efficiency and uniformity, making it possible to integrate over 100-mode MZI meshes in a whole wafer. There is no doubt to witness future generations of ZENs of larger scales and higher performance, and it would be exciting to eventually reveal the answer to the question of whether the PNNs can truly be useful in AI as a widely adopted hardware.

Moreover, from a long-term point of view, the TFLN platform can offer more distinguished characteristics for PNN development other than its high EO tuning efficiency shown here. For example, the high optical nonlinearity of TFLN combined with periodic poling technique makes it possible to incorporate optical nonlinear operations into the PNNs [32]. Furthermore, the capability of monolithic integration of passive and rare-earth-ion-doped active TFLN substrates allows to generate on-chip laser sources as well as light amplification for compensating optical loss [33–35]. The functionalities are yet to be explored for boosting the PNNs to an unprecedented level of computational power in the future.

## Acknowledgments

**Funding.** National Key R&D Program of China (2019YFA0705000, 2022YFA1404600, 2022YFA1205100), National Natural Science Foundation of China (Grant Nos. 12334014, 12192251, 12134001, 12304418, 12274130, 12004116, 12274133), Science and Technology Commission of Shanghai Municipality (NO.21DZ1101500), Shanghai Municipal Science and Technology Major Project (Grant No.2019SHZDZX01), Innovation Program for Quantum Science and Technology (2021ZD0301403).

**Disclosures.** The authors declare no conflicts of interest.

**Data availability.** Data underlying the results presented in this paper are not publicly available at this time but may be obtained from the authors upon reasonable request.

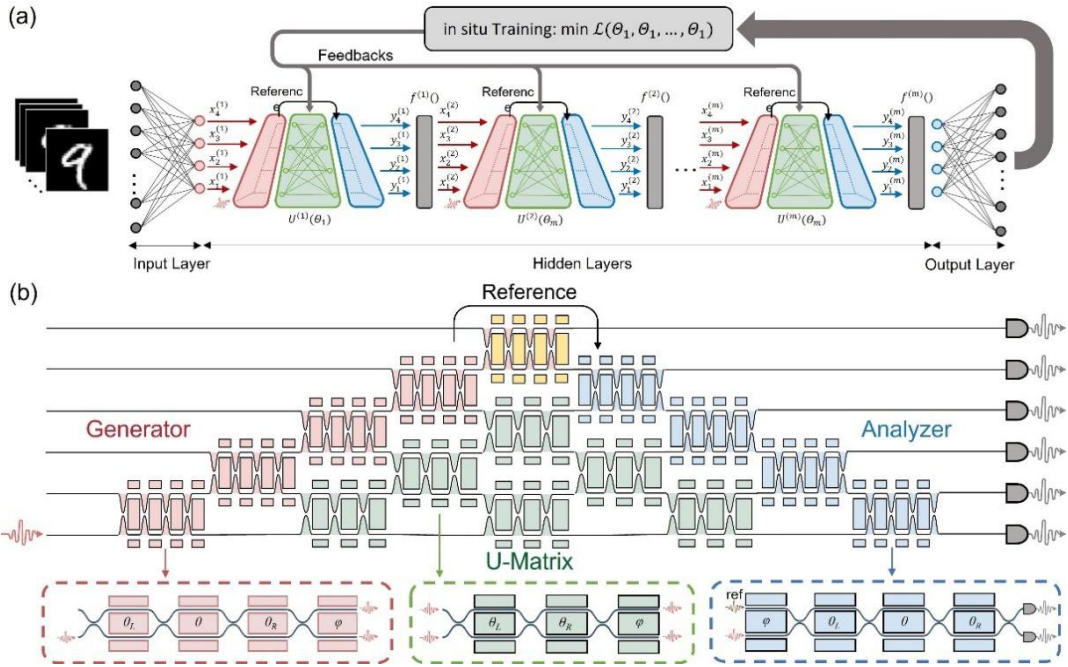

Fig. 1. (a) The architecture of a typical hybrid PNN consists of cascaded programmable unitary MVM operations assisted by electronic hardware to provide digital nonlinear functions as well as input and output layers on both ends. (b) The architecture of ZEN-1 comprises two 1x5 photonic meshes on each end of a programmable 4x4 triangle MZI mesh.

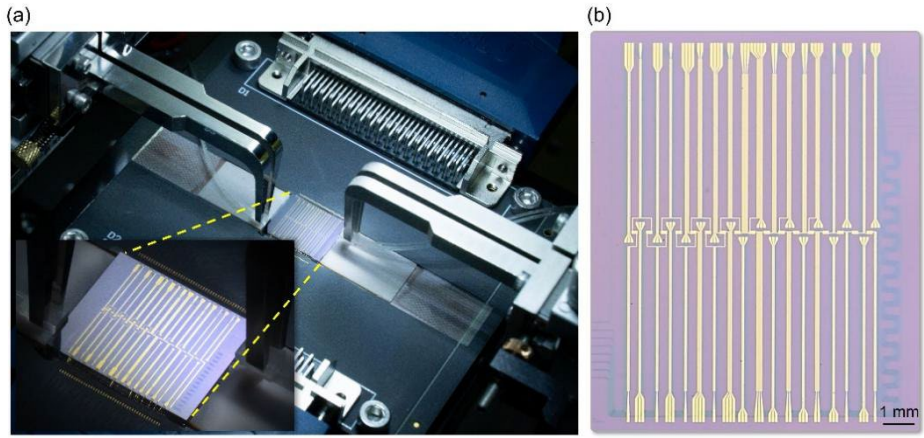

Fig. 2. (a) The photograph of the ZEN-1 was captured with a digital camera, lights are coupled in and out of the chip utilizing fiber arrays (FAs), and micro-electrodes of the chip are wire-bonded to a printed circuit board (PCB). (b) The microscope picture of ZEN-1, which consists 54 phase shifters.

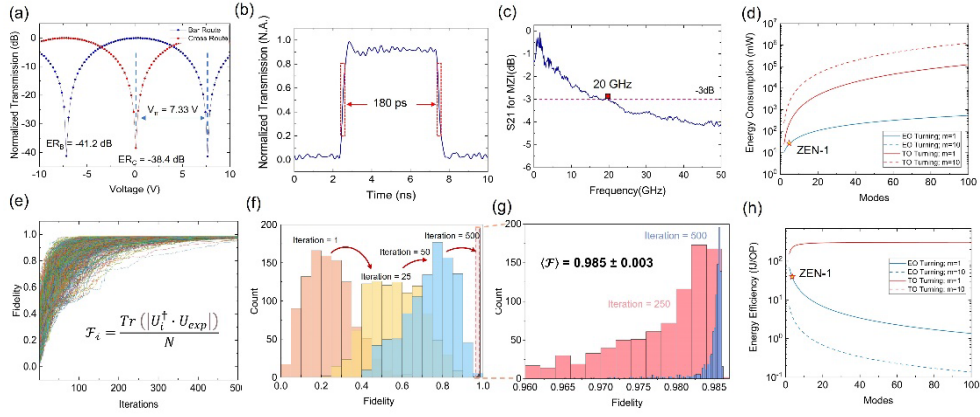

Fig. 3. (a) The measured EO response of a representative MZI in ZEN-1 for both bar route and cross route. (b) The optical transmission of the MZI as a function of time when applying a 0.1 GHz square-wave voltage, indicates a rise/fall time of 180ps. (c) The measured EO S21 curve of the MZI, indicates a bandwidth of 20 GHz. (d) Plot of total power consumption of PNNs that use EO/TO turning with m=1 and m=10 versus the number of modes. (e) Fidelities of ZEN-1 aiming at 1000 randomly generated unitary matrices during in-situ training. (f) Histogram of the 1000 fidelities after 1, 25, 50, and 500 iterations. (g) Zoomed in histogram of the 1000 fidelities after 250 and 500 iterations. (h) Plot of the power consumption for a single operation of PNNs that use EO/TO turning with m=1 and m=10 versus the number of modes.

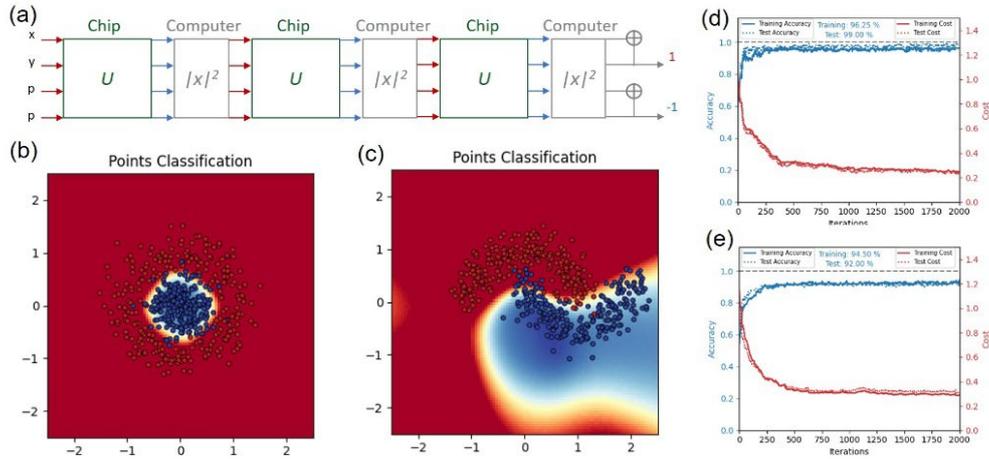

Fig. 4. (a) Schematic illustration of the architecture used to classify two labeled noisy synthetic datasets - the Circle and the Moons. (b) and (c) The decision maps of the two datasets after 2000 iterations. (d) and (e) In-situ training curves of the Circle and the Moons datasets.

# Supplemental Materials: Photonic Neural Network Fabricated on Thin Film Lithium Niobate for High-Fidelity and Power-Efficient Matrix Computation

Yong Zheng[1,2,†], Rongbo Wu[2,8,†], Yuan Ren[1,2], Rui Bao[1,2], Jian Liu[1,2], Yu Ma[3], Min Wang[2] and Ya Cheng[1,2,3,4,5,6,7]*

[1]State Key Laboratory of Precision Spectroscopy, East China Normal University, Shanghai 200062, China

[2]The Extreme Optoelectromechanics Laboratory (XXL), School of Physics and Electronic Science, East China Normal University, Shanghai 200241, China

[3]State Key Laboratory of High Field Laser Physics and CAS Center for Excellence in Ultra-intense Laser Science, Shanghai Institute of Optics and Fine Mechanics (SIOM), Chinese Academy of Sciences (CAS), Shanghai 201800, China

[4]Collaborative Innovation Center of Extreme Optics, Shanxi University, Taiyuan 030006, China.

[5]Collaborative Innovation Center of Light Manipulations and Applications, Shandong Normal University, Jinan 250358, China

[6]Shanghai Research Center for Quantum Sciences, Shanghai 201315, China

[7]Hefei National Laboratory, Hefei 230088, China

## Fabrication details

Our device is fabricated using a home-developed technology based on femtosecond laser direct write of Cr hard mask followed by chemo-mechanical polish (CMP) for transferring the mask patterns to the underneath TFLN, which consists of a 500-nm-thick TFLN layer bonded to a buried $SiO_2$ layer on a 500-μm-thick silicon support (NANOLN). The fabrication technical, known as PLACE [1] to the TFLN photonics community, is described with its process flow as shown in Fig. S1 with the following steps: (1) deposition of a 200 nm thick chromium (Cr) film on a commercial 4-inch LNOI wafer using magnetron sputtering; (2) patterning the pre-designed photonic structures on the Cr thin film using a femtosecond laser direct writing system; (3) transferring the pattern to the TFLN using CMP, which gives rise to a surface roughness of 0.1 nm, and in turn an extremely low propagation loss of 0.027 dB/cm [1]; (4) depositing a 1.5 μm thick layer of $SiO_2$ on the wafer as the waveguide cladding layer; (5) depositing a chromium-gold-titanium (Cr-Au-Ti) layer on the wafer with a thickness of 10 nm, 500 nm, and 100 nm, respectively, using magnetron sputtering; (6)

patterning Ti film into the mask for producing the gold electrodes using a femtosecond laser direct writing system; (7) wet-etching of gold using the Ti as the mask; and (8) removing the Ti mask as well as the excessive Cr adhesive layer on the bottom using wet chemical etching.

## Experimental setup details

To ensure the best EO modulation efficiency, we used a polarization controller (Polarization Synthesizer, PSY-201, General Photonics Corp., Chino, California, USA) to generate pure transverse-electric (TE) polarization of the input light at a wavelength of 1550 nm provided by a continuous-wave laser (CTL 1550, TOPTICA Photonics Inc., Farmington, New York, USA). We coupled light in and out of the device through fiber arrays and finally converted it into electronic signals using photon detector arrays (APD-2M-A-100K, Luster, Beijing, China). The electronic signals were recorded and processed by a synchronous analog output and data-acquisition system (PXIe-6739, National Instruments Corp., Austin, Texas, USA). Additionally, this system generated a 54-channel feedback signal for optimizing the chip configurations.

## Principle and design of the perfect MZIs

Foundry technologies often suffer inevitable fabrication errors, especially for delicate elements such as directional couplers (DCs) [2]. Likewise, for the DCs manufactured using PLACE technology, fabrication errors typically lead to a 10% fluctuation in the split ratio, making it challenging to construct MZIs with high extinction ratios (ERs) for high-fidelity MZI meshes. To address this issue, we add another 1 or 2 phase shifters (PSs) and DCs to build cascaded 2-MZI or 3-MZI units which function as the "perfect" MZIs as proposed by Miller [3]. The architectures of these perfect MZIs are shown in Figure S2(a). Here, we analyze the working principle of the architectures by considering their transmission matrices.

The transmission matrix of a single DC with a splitting ratio of $R$ can be expressed as:

$$T_{DC}(R) = \begin{bmatrix} \sqrt{R} & i\sqrt{1-R} \\ i\sqrt{1-R} & \sqrt{R} \end{bmatrix}, \quad (S1)$$

the transmission matrix of a single PS with a relative phase difference $\theta$ can be expressed as:

$$T_{PS}(\theta) = \begin{bmatrix} e^{i\frac{\theta}{2}} & 0 \\ 0 & e^{-i\frac{\theta}{2}} \end{bmatrix}. \quad (S2)$$

Thus, the transmission matrices for a traditional MZI, a cascaded 3-MZI unit and a cascaded 2-MZI unit as shown in Figure S2 (a) can be derived as follows:

$$T_{MZI} = T_{DC}(R_R) \cdot T_{PS}(\theta) \cdot T_{DC}(R_L); \quad (S3)$$

$$T_{3-MZI} = T_{DC}(R_{R1}) \cdot T_{PS}(\theta_R) \cdot T_{DC}(R_{R2}) \cdot T_{PS}(\theta) \cdot T_{DC}(R_{L2}) \cdot T_{PS}(\theta_L) \cdot T_{DC}(R_{L1})$$
$$= T_{MZI-L}(\theta_L) \cdot T_{PS}(\theta) \cdot T_{MZI-R}(\theta_R); \quad (S4)$$

$$T_{2-MZI} = T_{DC}(R_R) \cdot T_{PS}(\theta_R) \cdot T_{DC}(R_M) \cdot T_{PS}(\theta_L) \cdot T_{DC}(R_L). \quad (S5)$$

The operational principle of the cascaded 3-MZI units takes into account the effects of the PSs and DCs on both sides as two ratio-adjustable splitters (DC$_{L1}$, DC$_{L2}$ and $\theta_L$ for BSL; DC$_{R1}$, DC$_{R2}$ and $\theta_R$ for BSR). Hence, to construct an ideal MZI, one must adjust $\theta_L$ and $\theta_R$ until the split ratios of BSL and BSR reach 50:50. The simplified algorithm is depicted below.

1. Inject a laser beam of stable power into either input port T or L and apply a triangle-wave voltage on the middle PS, then monitor the extinction ratios (ERs) of the modulated light power from the two output ports (R and B).

2. Adjust $\theta_L$ to balance the ERs to ensure a 50:50 split ratio for the BSL.

3. Adjust $\theta_R$ to maximize the ERs to ensure a 50:50 split ratio for the BSR.

The EO response of a cascaded 2-MZI unit is more complicated since it involves simultaneously turning of both the PSs ($\theta_L$ and $\theta_R$) to achieve the predetermined split ratios. The calibration can be achieved by considering DC$_M$, DC$_R$ and $\theta_R$ as a tunable splitter BSR. When the split ratio of BSR equals the split ratio of the DC$_L$, the ER of the bar route (i.e. light transmitted from T/L to R/B) will reach its maximum. Conversely, when the split ratio of BSR compensates the split ratio of the DC$_L$, the ER of the cross route (i.e. light transmitted from T/L to B/R) will reach its maximum. Figure S2 (b) and (c) show the 3D plots of the ERs of the cross/bar route as the functions of $\theta_L$ and $\theta_R$. Evidently, when all split ratios of DCs are set as 30:70, both the routes can achieve high ERs regardless of the large fluctuations of split ratios of DCs.

## Working principles of the generator and the analyzer

A generator is a 1×N MZI mesh as shown in Figure S3 (a), which can generate optical modes for generating the complex vectors in any N-dimensional space with a single coherent input beam [4]. The processes of generating any predetermined complex vectors using such a structure can be described as follows.

Assuming a complex vector $|x\rangle = (x_1, x_2, x_3, x_4, x_5)^T$ and consider the last MZI in the generator, the transmission matrix can be expressed as follow for the light beam injected from the input port on the left-handed side and measured at the output port at the right-handed side:

$$\begin{bmatrix} x_5 \\ x_4 \end{bmatrix} = i \begin{bmatrix} e^{i\frac{\phi_4}{2}}\sin\frac{\theta_4}{2} & e^{i\frac{\phi_4}{2}}\cos\frac{\theta_4}{2} \\ e^{-i\frac{\phi_4}{2}}\cos\frac{\theta_4}{2} & -e^{-i\frac{\phi_4}{2}}\sin\frac{\theta_4}{2} \end{bmatrix} \begin{bmatrix} 0 \\ a \end{bmatrix} = a \begin{bmatrix} e^{i\frac{\phi_4}{2}}\cos\frac{\theta_4}{2} \\ -e^{-i\frac{\phi_4}{2}}\sin\frac{\theta_4}{2} \end{bmatrix}. \quad (S6)$$

Thus, $\phi_4$, $\theta_4$ and $a$ can be determined. Similarly, $\phi_3$, $\theta_3$ and $b$ can be determined as:

$$\begin{bmatrix} a \\ x_3 \end{bmatrix} = i \begin{bmatrix} e^{i\frac{\phi_3}{2}} \sin\frac{\theta_3}{2} & e^{i\frac{\phi_3}{2}} \cos\frac{\theta_3}{2} \\ e^{-i\frac{\phi_3}{2}} \cos\frac{\theta_3}{2} & -e^{-i\frac{\phi_3}{2}} \sin\frac{\theta_3}{2} \end{bmatrix} \begin{bmatrix} 0 \\ b \end{bmatrix}. \tag{S7}$$

Following the same principle, all the parameters in the generator can be determined.

Theoretically, an analyzer can be considered as a generator operated in reverse, i.e., turning all PSs in the analyzer to combine all input light beams into one single port, and after that the output complex vector $|y\rangle = (y_1, y_2, y_3, y_4, y_5)^T$ can be extracted from the final settings of all PSs in the generator [4]. In our experiment, we extract $|y\rangle$ by setting $\theta = \pi/2$ and $\varphi = 0, \pi/2, \pi, 3\pi/2$ in each MZI in generators and measure the output powers $p_0$, $p_{\pi/2}$, $p_\pi$, $p_{3\pi/2}$ and compute the relative phase as $arctan((p_{3\pi/2} - p_{\pi/2})/(p_\pi - p_0))$ as schematically shown in Figure S4 (b).

## Details of the Calibration processes

Before the calibration process, light signals will be routed randomly into a large number of optical paths of the MZI mesh due to the inevitable initial phase errors in every fabricated waveguides. The calibration processes of our device are schematically illustrated in Figure S4. First, we couple a laser beam into the top left input (L6) of the mesh and measure the transmissions at the right ports (R6-R1), as shown in Figure S4(a). This procedure calibrates the internal PSs of all MZIs on the right-handed side of the mesh, i.e., all the MZIs in the analyzer. Then, we implement the similar procedure by injecting the laser beam into the mesh in the opposite direction, i.e., this time the input light is coupled into the top right input (R6) of the mesh and then the transmissions at the left ports (L6-L1) are measured. The internal PSs in the middle triangle MZI mesh are calibrated based on the topology and order of the MZIs using the same procedures in our previous work [5], as shown in Figure S4 (b).

The external PSs of the generator and analyzer are calibrated using several "meta-MZIs." Figure S4 shows that a "meta-MZI" consists of two MZIs arranged in columns, which are programmed to act as the 50-50 beam splitter ($\theta = \pi/2$). This sub-circuit can be regarded as an effective MZI, where the relative phase between two external PSs is equivalent to the setting of the internal PS in a discrete device. We sweep the voltage applied on one of the two external PSs, measure the output transmission, and adjust the voltage applied to the other until the output transmission curve fulfills a standard sinusoidal waveform with zero initial phase.

## Stochastic Optimization Processes

We used a stochastic optimization method that performs gradient descent on average and converges to a local minimum. An iteration of the optimization process can be summarized as follows:

1. Perturb all weights $\Theta$ (all voltages applied to the PSs in the SU(4) in ZEN-1) towards a random direction $\Delta$ in search space, i.e., $\Theta \rightarrow \Theta + \Delta = \Theta + [\delta_1, \delta_2, \ldots, \delta_N]$, calculate the cost $\mathcal{L}(\Theta + \Delta)$.
2. Change all weights in the opposite direction, i.e., $\Theta \rightarrow \Theta - \Delta = \Theta - [\delta_1, \delta_2, \ldots, \delta_N]$, and calculate the cost $\mathcal{L}(\Theta - \Delta)$.
3. Calculate the directional derivative $\nabla_\Delta \mathcal{L}(\Theta) = (\mathcal{L}(\Theta + \Delta) - \mathcal{L}(\Theta - \Delta))/2\|\Delta\|$.
4. Update all weights to $\Theta \rightarrow \Theta - \eta \nabla_\Delta \mathcal{L}(\Theta)\Delta$ with a learning rate $\eta$ chosen as a hyperparameter.

## Classification of Iris flowers and MNIST handwriting digits

Our second application of PNN involves the classification of Iris flowers. The objective is to classify an Iris flower into one of three subspecies (setosa, versicolor, and virginia) based on four parameters: the length and width of the petals and sepals. We utilized a three-layer architecture, as shown in Fig. S5(a). The train–test split ratio was chosen as 80%:20% (120 train samples, 30 test samples). We sent the training set into the system and implemented the in-situ training using a cross-entropy cost function.

The training curves for 5000 iterations shown in Fig. S5(b) indicate a train/test accuracy of 88.33%/90.0%, and the confusion matrices of the train/test datasets are shown in Fig. S5(c).

Our final demonstration of PNN involves using ZEN-1 to classify handwritten digits in the MNIST dataset. The architecture of our model is shown in Fig. S6 (a). We first mapped the input 28×28 grayscale images to 784×1 vectors and fed them into our system. The model consists of one 784×4 input layer, followed with three 4×4 hidden layers connected with ReLU digital nonlinear functions, and one 4×10 output layer that maps the 4x1 output vector to 10 classes which represent the digits from 0 to 9. The hidden real-valued matrices ($M$) layers can be decomposed as $M = U\Sigma V^{\dagger}$ through singular value decomposition (SVD), in which the unitary matrices $U$ and $V^{\dagger}$ are implemented on ZEN-1 whereas the input and output layers and the diagonal matrix $\Sigma$ are executed electrically. Instead of performing in-situ training, we created a simulation model based on TanserFlow and completed the training using a computer. This allowed us to obtain all matrix elements of the three hidden layers and then train ZEN-1 aiming at these matrices. We then fed all 60000 training and 10000 testing samples of the MNIST dataset into our system. We ran 70000 inferences and achieved a train/test accuracy of 88.5%/89.3%. The confusion matrices of the train/test datasets are shown in Fig. S6 (b).

A. B. Miller, "Experimentally realized in situ backpropagation for deep learning in photonic neural networks," Science **380**, 398–404 (2023).

[5] Y. Zheng, H. Zhong, H. Zhang, L. Song, J. Liu, Y. Liang, Z. Liu, J. Chen, J. Zhou, Z. Fang, M. Wang, L. Li, R. Wu and Y. Cheng, "Electro-optically programmable photonic circuits enabled by wafer-scale integration on thin-film lithium niobite," Phys. Rev. Res. **5**, 1–9 (2023).

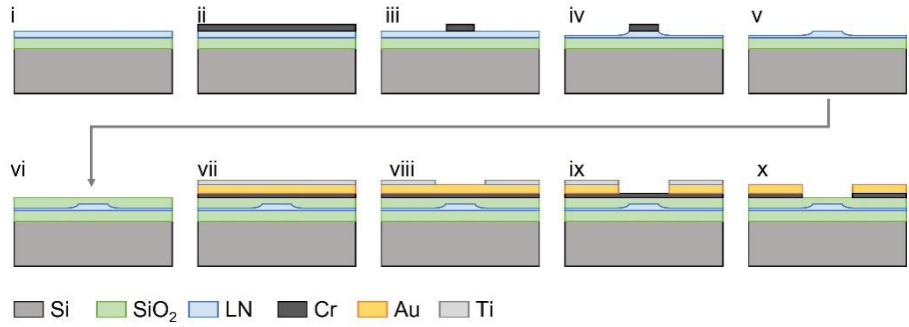

Fig. S1. Schematic diagram of the fabrication flows of PLACE technology.

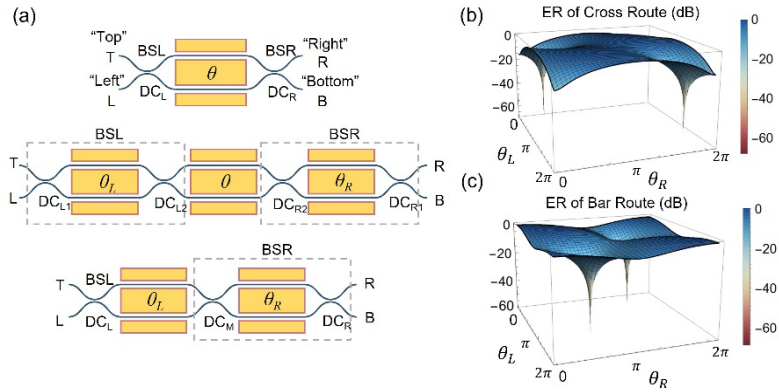

Fig. S2. (a) The architectures of 3-cascaded and 2-cascaded MZIs for performing "perfect" MZIs with ultimate high extinction ratios. (b) and (c) The 3D plots of the ERs of the cross/bar route as functions of $\theta_L$ and $\theta_R$, with all split ratios of DCs set as 30:70.

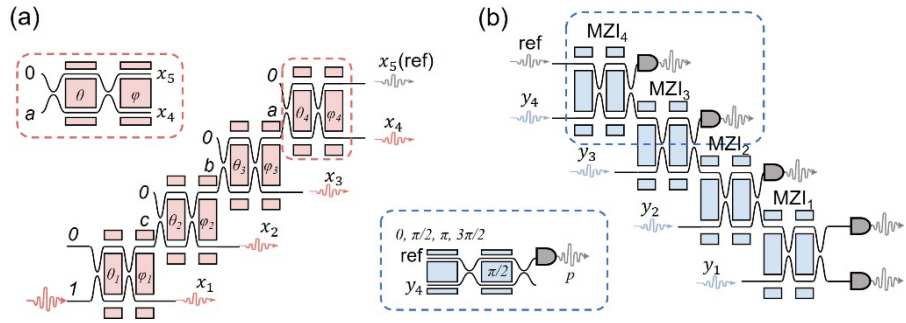

Fig. S3. (a) Schematic of the generator. (a) Schematic of the analyzer consists of cascaded phase-diversity optical homodyne detectors.

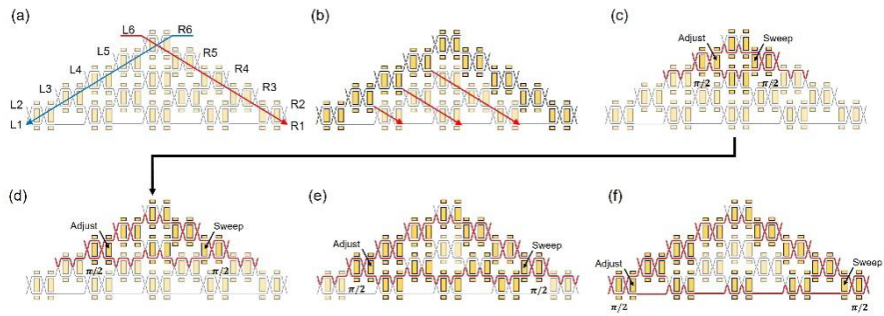

Fig. S4. (a) and (b) Calibration procedure for internal phase shifters in ZEN-1. (c) – (f) Calibration procedure for external phase shifters in generator and analyzer.

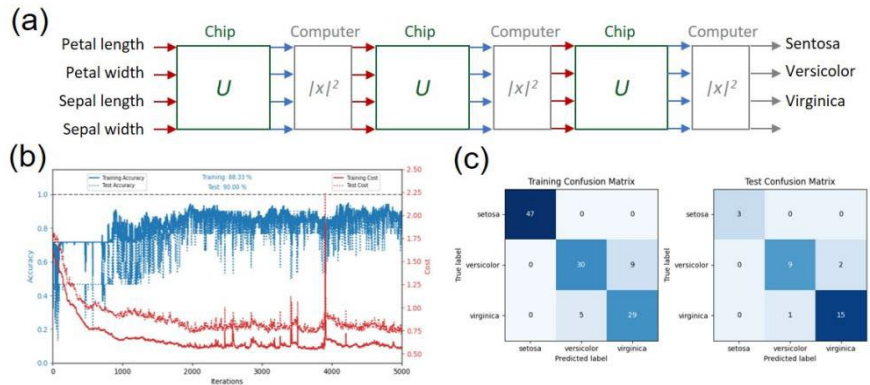

Fig. S5. (a) Schematic illustration of the architecture utilized for classifying an Iris flower into one of three subspecies. (b) The training curves for 5000 iterations. (c) The confusion matrices of the train/test datasets after 5000 iterations.

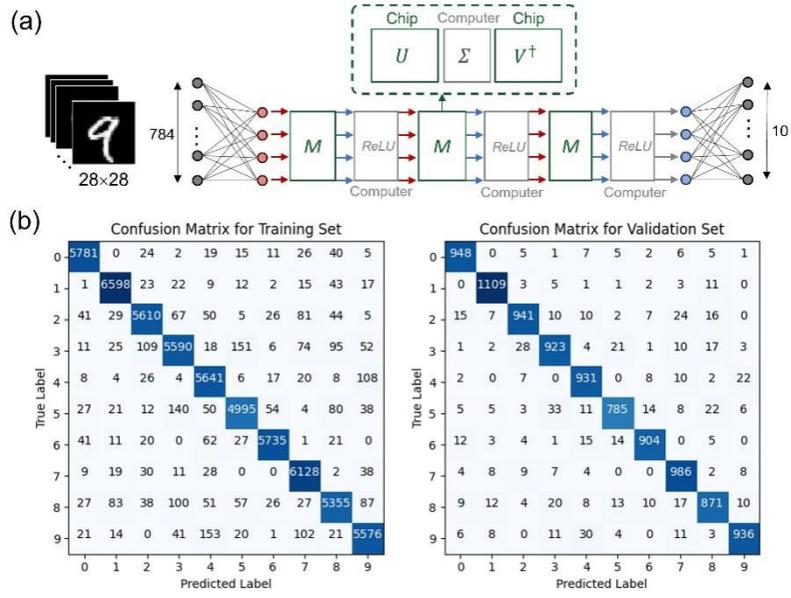

Fig. S6. (a) Schematic illustration of the architecture utilized for classifying handwritten digits in the MNIST dataset. (b) The confusion matrices for 60000 training and 10000 testing samples of the MNIST dataset inferenced on our hybrid ANN model.